\documentstyle[preprint,aps,psfig]{revtex} 
\begin{document}
\draft

\title{Static magneto-polarizability of cylindrical nanostructures}

\author{St\'{e}phane Pleutin and Alexander Ovchinnikov}
\address{Max-Planck-Institut f\"ur Physik Komplexer Systeme, N\"othnitzer
Stra\ss e 38, D-01187 Dresden}

\date{\today}
\maketitle

\begin{abstract}
The static polarizability of cylindrical systems is shown to have a strong
dependence on a uniform magnetic field applied parallel to the tube
axis. This dependence is demonstrated by performing exact numerical diagonalizations of
simple cylinders (rolled square lattices), armchair and zig-zag carbon
nanotubes (rolled honeycomb lattices) for different electron-fillings. At low
temperature, the polarizability as function of the magnetic field has a discontinuous character where
plateau-like region are separated by sudden jumps or peaks. A one to one
correspondence is pointed out between each discontinuity of the polarizability
and the magnetic-field induced cross-over between the ground state and the
first excited state. Our results suggest the possibility to use
measurements of the static polarizability under magnetic field to get
important informations about excited states of cylindrical systems such as carbon nanotubes.
\end{abstract}

\vskip2pc
\narrowtext
\section{Introduction}
Cylindrical like systems, as nanotubes of carbon or one dimensional stacks
of ring organic molecules, show, in the absence of time-reverse symmetry breaking
perturbations, two energetically degenerated families of one-electron
states. One is for electrons moving in clockwise direction, the other is for electrons moving
in counterclockwise direction. A magnetic field applied along the cylindrical
axis breaks time-reverse symmetry. The degeneracies are then lifted and a diamagnetic current is
induced. If one considers the behaviour of the energies of the many-particle
states, one finds lot of level crossing induced by the magnetic field. In other
words, accidental degeneracies are created at some values of the external
magnetic field. This means also that the ground state changes when the magnetic
field varies.

Very recently \cite{fulde} the study of the polarisability of
cylindrical systems under magnetic field was suggested as a possible probe to
analyse their electronic structures. The underlying mechanism of this
proposed new spectroscopy is quite simple. The magnetic field induces level
crossings as already mentioned. As a consequence of that, changes of the ground state occur which at some
fields is degenerate. Applying in addition an electric field, two kinds of effect are
expected near those magnetic-field-induced accidental degeneracies:
(i) at the crossing points, a linear Stark effect may occur if there exists a
non vanishing
matrix element of the perturbation between the two crossing states, resulting
in divergencies in the polarisability, (ii) when the matrix element between
these
two states is vanishing, the system shows a quadratic Stark effect but the difference between the
quadratic Stark coefficients of the two states involved creates discontinuities in the
response function. A careful analysis of the linear electric susceptibility
should then provide important informations about properties of excited states
such as their energies and symmetries. A few years ago it was already pointing out that a magnetic field
should have pronounced effects on the polarisability but for quite different
materials and purposes, i.e. in \cite{efetov,bouchiat,blanter} small metallic
particles (rings, disks or spheres) were considered in connection with weak
localization. In particular, it was found that the polarizability should be
greater in magnetic field than in zero field because of the disappearance of
weak localization. This theoretical prediction was observed very recently in
the ac polarizability of mesoscopic rings \cite{deblock}. 

Square lattices and honeycomb lattices (nanotubes) rolled-up into cylinders with uniform electric and
magnetic fields both applied along the cylindrical axis were studied in
\cite{fulde}. With this configuration full quantum calculations were not
possible for large systems. Instead, since the electric field creates a smoothly varying
potential across the cylinder, a semi-classical expression for the dielectric
function was used. With this approximate approach, the discontinuities of the
polarisability were well observed - resulting in extraordinary rich structures - however,
the effects of possible linear Stark effect were not described.

The main purpose of this work is to establish and extent on the basis of full quantum
calculations the ideas and concepts discussed in \cite{fulde} using
semi-classical calculations. For that purpose, the very same cylindrical systems are
considered, i.e. square and honeycomb lattices, but with
the modification that the electric field is applied perpendicularly to the
cylindrical axis (cf figure 1). The net advantage of this choice is to allow
for a separation of variable, i.e. the two degrees of freedom corresponding to the
motion of the electrons along the circumference and the cylindrical axis of
the cylinder can be treated separately. The exact calculation
of the polarisability (i.e. by fully quantum treatment) is then possible even for very
large systems. The first conclusions of \cite{fulde} are confirmed and the
appearance of linear Stark effect are well identified. Additionally, the effects
of the shape of a system and of the Zeeman interaction on the dielectric
function are discussed as
well as differences occurring in the magnetic field dependent spectrum of armchair and zig-zag
nanotubes. Finally perturbative calculations of the dielectric response are done. They are in perfect agreement
with the exact results as long as we remain in the linear regime.

Before proceeding further we want to stress that all the necessary conditions are
available nowadays to realize the experiments we are proposing. On one hand very accurate
measurements of the polarisability at very low temperatures are possible \cite{deblock,glass}. A
spectacular example was given by the recent observations of a strong magnetic-field
dependence of the polarisability of multicomponent glasses in the mK regime 
\cite{glass}. On the other hand,  systems with diameters in the mesoscopic
range are required in order to realize the experiment; however large circumference nanotubes
are routinely produce today\cite{saito-book} and could be first candidates of
interest.

\section{Rolled square lattices}
We consider first cylinders in form of rolled square lattices. As already mentioned in
the introduction, two uniform fields are
applied to those systems: a magnetic field
$H$, parallel to the cylindrical axis, and an electric field $E$,
perpendicular to it (cf figure 1). In this work we are concerned only with
orbital magnetism. Therefore we will neglect the spin  of the electrons. The
dynamic of the spinless fermions are described in terms of the following standard tight-binding model.

\begin{equation}
\label{huckel}
\hat{H}=t\sum_{n,m}(a^{\dagger}_{n+1,m}a_{n,m}e^{i\frac{2\pi}{N}\phi}+h.c.)+t_{p}\sum_{n,m}(a^{\dagger}_{n,m+1}a_{n,m}+h.c.)+v\sum_{n,m}\cos(\frac{2\pi}{N}n)a^{\dagger}_{n,m}a_{n,m}
\end{equation} Different sites are labelled by the indices $n$ along the circumference running from $1$ to $N$,
and $m$ along the cylindrical axis running from $1$ to
$M$. The total number of sites of the cylinder is thus given by $NM$. $a^{\dagger}_{n,m}$ ($a_{n,m}$) is the creation (annihilation) operator of
a spinless electron on site $(n,m)$. $t$ and $t_p$ are the nearest-neighbour hopping integrals, $\phi$
is the magnetic flux in unit of the elementary flux $\phi_0=\frac{\hbar c}{e}$
and $v$ denotes the potential related to the electric field, i.e.
$v=eRE$, with $R$ being the radius of the cylinder.

The field $E$ is supposed to be small enough so that we are in the linear
response regime. The magnetic flux is proportional to the magnetic field and the section area of the cylinder
\begin{equation}
\phi=\frac{N^{2}a^{2}H}{8 \pi^{2}}
\end{equation} where $a$ is the lattice constant on the circumference. We are
interested in systems of a mesoscopic size along the
circumference. Typically, $R$ should be in the range of several tens of
nanometres. Then the corresponding flux quantum is of order several tens of
tesla which is nowadays reachable experimentally.

The effects of the Zeeman interaction are not considered in this work. With
this interaction, if one includes also the spin-orbital coupling, changes can arise in the
calculated polarizability especially at high magnetic field or low electronic
density as briefly discussed below. The effects of the orbital magnetism
discussed in this work are expected to be in any case predominant. Nevertheless, for practical
purpose, the effects of the Zeeman and spin-orbital interactions should be
also incorporated.

Moreover the electron-electron interaction terms are not explicitely introduced in
this work. Instead they are supposed to be included in the effective
one-electron parameters of our model in the spirit of the Fermi-liquid theory of Landau. An explicit treatment of these interactions which
lead to screening of the electric field 
could produce important qualitative changes especially in the strong
coupling limit as it was shown for ring systems \cite{kusmartsev}.

Without electric field, i.e. for $v=0$, the spectrum of (\ref{huckel}) is given by

\begin{equation}
\label{spectrum}
\epsilon_{p,q}=2t\cos(\frac{2\pi}{N}(p+\phi))+2t_p\cos(\frac{\pi}{M+1}q)
\end{equation} with $-N/2 \leq p \leq N/2-1$ and $ 1 \leq q \leq M$. We have applied open
boundary conditions at the ends of the cylinder. It has to be associated with
the one-electron wave functions
\begin{equation}
\label{wavefunction}
|\Psi_{p,q}> = \sqrt{\frac{2}{N(M+1)}}
\sum_{n=1}^{N}\sum_{m=1}^{M}e^{i\frac{2\pi}{N}pn}\sin(\frac{\pi}{M+1}qm
)a^{\dagger}_{n,m}|0>
\end{equation} where $|0>$ is the vacuum.

At zero magnetic field, i.e for $\phi=0$, the spectrum is two fold degenerate,
$\epsilon_{p,q}=\epsilon_{-p,q}$, except for the states with $p=0$ and
$p=-N/2$. Adopting the convention that states with positive
$p$ are for electrons running in clockwise direction then states
with negative $p$ describe electrons moving in counterclockwise
direction.

A finite magnetic
field ($\phi \ne 0$), breaks time-reversal symmetry and implies lifting the
two-fold degeneracy and inducing a diamagnetic current. As a consequence, the
energy spacings of the many-electron states are continuously changing with increasing magnetic
field. This is shown in Fig 2a for the lowest eigenstates.

For a finite electric field ($v \ne 0$), it is not possible anymore to solve
analytically model Hamiltonian
(\ref{huckel}). However, for the configuration shown in Fig 1, we can
treat separately the variables $n$ and $m$. This reduces the study to a one
dimensional Hamiltonian for a ring in an applied electric field

\begin{equation}
\label{harper}
\hat{H}_R=t\sum_{n}(a^{\dagger}_{n+1}a_{n}e^{i\frac{2\pi}{N}\phi}+h.c.)+v\sum_{n}\cos(\frac{2\pi}{N}n)a^{\dagger}_{n}a_{n}
\end{equation} This is Harper's model which has
been extensively used in very different
contexts of condensed matter physics and which can be treated. It should be
noticed that, this model, i.e, a ring
placed in an uniform electric field - is similar to a rectangular lattice
with hopping integrals given by $t$ and $v/2$ and threaded by
a magnetic field with a flux given by $\phi=\frac{2\pi}{N}$.

In the following, we shall assume that $t=t_p$.
The relative value of these two transfer integrals has considerable influences on the
dielectric function but we leave this study to future considerations.

Once the spectrum is known, we calculate the induced
dipole moment $D$, as function of the magnetic field and temperature $T$ as follows.

\begin{equation}
\label{dipole}
D(T,\phi)=\frac{Tr \hat{d}e^{-\beta(\hat{H}-\mu)}}{Tre^{-\beta(\hat{H}-\mu)}}
\end{equation}where, as usual, $\beta=\frac{1}{k_B T}$ and $\hat{d}$ is the
dipole operator
\begin{equation}
\label{dipoleoperator}
\hat{d}= eR\sum_{n,m}\cos(\frac{2\pi}{N}n)a^{\dagger}_{n,m}a_{n,m}
\end{equation}

We show in figure 2b the magnetic field dependence of the polarisability for
a cylinder with $N=101$ and $M=100$ and a very few electrons on it, $N_e=100$
(which corresponds to a band filling of only $1 \%$). In this example, the electric field is such that
$v=10^{-3}t$ and the temperature is $k_bT=10^{-5}t$. We choose this example
because it shows very clearly the main behaviours of the dielectric response. In particular we choose $N=101$ because this gives an
illustration of the signature of the linear Stark effect. 

First of all the induced
dipole moment as function of the magnetic field is periodic with a period of
$\phi_0$ and is symmetric with respect to $\phi_0 /2$; these symmetries are
already apparent in the spectrum (\ref{spectrum}). Second, the induced
dipole moment shows clearly two main characteristics: (i) we can
notice the presence of small peaks at $\phi=0, 1/2, 1$, (ii) the induced dipole moment is a discontinuous function showing
several jumps separating plateau-like sections. Note that there is a slight curvature in the whole
spectrum, which is related to the persistent current induced by the
magnetic field. 

As noticed before, a magnetic field induces crossing between the
energies of the ground state and the first excited state (figure 2a). As it
can be seen in figures 2, there is a one to
one correspondence between level crossing at zero electric field and each
kink in the induced dipole moment. 

At each crossing, the ground state of the system changes. These two states
which are crossing respond differently to an
applied electric field. Both produce a quadratic Stark effect but generally of
different size. This explains why the
induced dipole moment is not a continuous function of the magnetic field. 

Near the crossing points, the response of the system will depend on whether or
not there is a nonzero matrix element of the electric field between the two states
involved. If the matrix element vanishes, the picture described above is
valid. On the contrary, if there
is interaction between those states, due to the degeneracy,
the response will become a linear (instead of quadratic) Stark effect resulting in peaks of the induced
dipole moment. More precisely, using the expression of the wave function
(\ref{wavefunction}), the matrix elements of the dipole operator can be
calculated
\begin{equation}
\label{selectionrules}
< \Psi_{p,q}|\hat{d} | \Psi_{p',q'}>=eR \delta_{p',p\pm1} \delta_{q',q}
\end{equation} With this equation, it is easy to see that linear Stark effect
could occur only for the following very particular values of the magnetic flux:
$\phi=0,1/2,1$. These are precisely the values for which one gets peaks in our first example (figure 2b).

It is worth notice that large enough Coulomb interaction could
change drastically the selection rules (\ref{selectionrules}). Appearance of
linear Stark effects for new values of the magnetic flux could then give a way
to quantify importance of electronic correlation effects.

At low enough temperature, one deals essentially with the spectroscopy of a few
levels around the Fermi level. Therefore it is not surprising that the induced
dipole moment for a well definite system, i.e., one which is well ordered, well oriented and of well
defined size ($N,M$), depends strongly on the electronic
density. This is clearly apparent from Figs. 3.a and 3.b. They are for the
very same system than before but for three different electron fillings
$N_e=100, 101$ and $102$. At such a low density, the relevant mean level
 spacing behaves as $\Delta E \approx 1/N^2$. Figure 3.a is for a
temperature lower than $\Delta E$ by one order of magnitude while Fig. 3.b is
for a temperature higher than $\Delta E$ by one order of magnitude. With
a typical value for $t$ ($t\approx 2 eV$) one can estimate a temperature of
$10^{-1}K$ for the spectrum 3.a and $10K$ for the spectrum 3.b. With these
figures we want to emphasize the unique sensitivity of the proposed measurements
and its corollary which is the necessity to work at very small temperature in order to
get the maximum informations. 

In Fig. 4, we show again the induced dipole moment for the same electric field
and temperature but for a bigger cylinder, $N=101$
and $M=1000$, and more electrons on it, $N_e=20000$ (electron density of $20
\%$). The response appears to be much
more complex but shows the same characteristics of peaks and plateau-like parts
separated by discontinuous jumps. This is more apparent in the inset which
shows a zoom of the spectrum at low magnetic fields.

Today it is possible to measure accurately very small variations in the real part of the dielectric
function \cite{glass}. Therefore, the dramatic magnetic field effects on the
static polarisability of mesoscopic cylindrical systems, discussed in this work,
could be measured and analysed. Several important informations about excited states could then be
obtained. First, the positions of the discontinuities
and peaks should give informations about the energies of the excited states. The nature of the
response - linear or quadratic Stark effect could be detected and therefore
should give information about the symmetry of the excited states. The
magnitude of the response should give also informations about the coupling
constants. Finally, the different curvatures observed in the whole spectrum
could give information related to the persistent current induced by the
magnetic field. 

Of course, as it was already mentioned, the model we are studied gives an
oversimplified view of the reality. In order to go to realistic
systems several other aspects remain to be clarified. The very important case
of electron-electron interactions will be the subjects of subsequent works. In the following, we
discuss briefly, first, possible influences of the shape of the
cylindrical cross-section and, second, the role plays by Zeeman interaction. For
practical purposes, it is certainly necessary to study in details both of these points.

{\it Influences of the shape. Case of elliptical cross-section}. Until now, we have
considered only cylindrical systems with circular cross-section. With this
particular shape, a uniform electric field creates the cosine potential
appearing in (\ref{huckel}). The electronic eigenstates of these cylindrical
systems are characterized by a pair of wave vectors, $k$ and $q$, for the
motion along the circumference and the cylindrical axis respectively. The
cosine potential of equation (\ref{huckel}) couples states
of different wave vectors, $k_1$, $q_1$ and $k_2$, $q_2$, in such a way that
the following selection rules are
fulfilled $\Delta k=k_1-k_2=\pm \frac{2\pi}{N}$ and $\Delta q=q_1-q_2=0$, as
already discussed above. However, there exist many cases
where the section of the
cylinder may have different shapes. One could think, for instance, of
one a dimensional stack of large organic polycyclic molecules which do not
form regular circles according to the well known hybridisation properties of
carbon atom. With different
shapes the cosine potential will be affected undergoing new selection
rules. These new selection rules could in turns influence substantially the
induced dipole moment. For illustration we
consider here the case of an elliptical cross-section and compare its response
with the one of a circular section.

For a general cross-section the dipole operator takes the following form

\begin{equation}
\label{dipoleoperatorellipse}
\hat{d}= e\sum_{n,m}R(n)\cos(\Theta(n))a^{\dagger}_{n,m}a_{n,m}
\end{equation}where $R(n)$ is the distance of the site $n$ to the cylindrical
axis and $\Theta(n)$ the corresponding polar angle with respect to some
arbitrary axis. The sites are supposed to be equally space. $R(n)$ and
$\Theta(n)$ are determined under this condition. The model (\ref{huckel}) is
then still valid and the electronic spectrum of the system without electric
field is still given by Eq. (\ref{spectrum}). 

An ellipse is characterized by two parameters, the major axis $2a$, and the
minor axis $2b$. We give an
example for an elliptical cylinder with $N=100$, $M=100$, $N_e=100$ and
$\frac{b}{a}=0.5$.

In Figs. \ref{ellipse}a and \ref{ellipse}b are reported the induced dipole moment for elliptical
and circular systems with $v=10^{-3}t$; Figs. \ref{ellipse}c and \ref{ellipse}d
present the same results but for $v=10^{-2}t$.

For an ellipse, there is additional coupling between $k$ states which do not
fulfilled the original selection rules $\Delta k=\pm \frac{2\pi}{N}$ and
$\Delta q=0$. At small
electric fields (linear regime) these additional couplings yield very smooth
changes only on the
shape of the induced dipole moment; as can be seen in Figs. \ref{ellipse}a and \ref{ellipse}b,
only the amplitudes are slightly modified. However, at higher electric fields
more dramatic changes appear (Figs. \ref{ellipse}c and \ref{ellipse}d). This is the case
in our example where one can notice, for instance, the appearance of a new
peak at $\phi \simeq 0.26$ for an elliptical section.

{\it Influences of the Zeeman interaction}. For realistic consideration
of fermions with spin $1/2$, the Zeeman interaction combined with the spin-orbital interaction
must be clarified. In this subsection we give only a first hint in that
direction.

The Zeeman Hamiltonian is given by

\begin{equation}
\label{zeeman}
\hat{H}_Z=g \mu_B\vec{S}\vec{B}
\end{equation}
where $\vec{S}$ is the total spin of the system, $\mu_B$, the Bohr magneton
and $g$, the Land\'e factor. Due to the effect of this interaction every
one-particle level will be split into a spin-up and spin-down component by a
term proportional to $\phi/N^2$. Therefore, by considering also the effects of
the spin-orbit
coupling, the whole spectrum could be changed: both, the positions of the
accidental degeneracies (discontinuities) could be shifted and the intensities of the induced
dipole moment could be modified. The importance of those changes can be
estimated by considering the ratio of the Zeeman energy $~\phi/N^2$ and,
$\Delta(n)$, the level spacing of the one-dimensional ring Hamiltonian
(\ref{harper})

\begin{equation}
\label{levelspacing}
\Delta(n)=4t\sin(\frac{2\pi}{N})\sin(\frac{2\pi}{N}(n+\phi+\frac{1}{2}))
\simeq \frac{4\pi}{N}\sin(\frac{2\pi}{N}n)
\end{equation}

The spacing is not an uniform function of $N$. It behaves as $1/N^2$ at the
bottom of the band, and as $1/N$ in the middle of the band. With this
consideration one may conclude that the Zeeman plus spin-orbit coupling can
become important in the case of (i) high magnetic fields or/and (ii) low
electronic density.

\section{Armchair and Zig-Zag carbon Nanotubes}

Within the class of cylindrical materials, carbon nanotubes are certainly among
the most interesting and fascinating. They are honeycomb lattices rolled into
cylinders \cite{lijima}. Part of their interests comes from their unique interplay
between geometry and electronic properties \cite{saito-book}. Indeed, a single-wall nanotube can be either metallic or
semiconducting
depending on its diameter and its chirality. This fact was recognized very soon
after their discovery using tight binding models
\cite{saito} and from first principle calculations \cite{hamada}. 

We consider in the following the two simplest kinds of nanotube: the so-called zig-zag
nanotubes (Fig. \ref{nanolattice}a), which are semiconductors (conductors) if the number of
unit cell $N$ is not (is) a multiple of 3, and the armchair nanotubes
(Fig. \ref{nanolattice}b),  which are always metallic \cite{saito-book}. They are the two kinds of nanotube having the highest symmetry.
Moreover they are the only two examples showing no chirality. Because of this
last characteristics, they can be considered as a kind of rolled square
lattice but with four
carbon atoms per unit cell; these units are
shown on the Figs. \ref{nanolattice} for both systems. In the presence of an
uniform magnetic field along the cylindrical axis, the spectrum is formally
similar for the two systems

\begin{equation}
\label{spectrumnano}
\epsilon_{p,q}=\pm (1+u_p \pm (u_pv_q)^{1/2})^{1/2}
\end{equation}where for zig-zag nanotubes
\begin{equation}
\label{zig-zag}
\begin{array}{c}
u_p=2(1+\cos(\frac{2\pi}{N}(p+\phi))),\quad \mbox{with} \quad p=1,..,N\\
\\
v_q=2(1+\cos(\frac{\pi}{M+1}q)),\quad \mbox{with} \quad q=1,..,M
\end{array}
\end{equation}and for armchair nanotubes
\begin{equation}
\label{armchair}
\begin{array}{c}
u_p=2(1+\cos(\frac{\pi}{M+1}p)),\quad \mbox{with} \quad p=1,..,M\\
\\
v_q=2(1+\cos(\frac{2\pi}{N}(q+\phi))),\quad \mbox{with} \quad q=1,..,N
\end{array}
\end{equation}The roles of $u_p$ and $v_q$ are just exchanged from one case to the
other. For calculating these spectra we have used the transformation from
a hexagonal lattice to rectangular lattice with four sites per unit-cell introduced in \cite{sigrist}.

With an electric field, the spectrum can no longer be obtained analytically
however - as for the case of rolled square lattices - since the systems
chosen have no chirality, it is still possible to treat
separately the variables along the circumference and the cylindrical axis. The
effective systems we have then to consider explicitely are rings of $N$ units,
containing each four carbon atoms, but where the coupling constants depend on the wave vector in the
cylindrical direction. Proceeding that way, it is then possible to study exactly
the response to an electric field even for very large systems.

In our calculations we neglect the two ends of the nanotubes which consist of
a "hemisphere" of a fullrene \cite{saito-book}. Since the scale along the
cylindrical axis is in our calculations much larger than the one along the diameter this
approximation should be justified (in reality ratio
as large as $10^5$ between these two characteristic scales are usual).

We present first, results for armchair and zig-zag nanotubes with $N=50$ and
$M=500$ at half-filling, $N_e=50000$, for a small electric field,
$v=10^{-3}t$, and low temperature, $k_BT=10^{-5}t$. For this choice of $N$,
the zig-zag nanotube is semiconductor. The figures \ref{armchairg}a and \ref{armchairg}b show the ground state and first
excited state energies and the induced dipole moment, respectively, as
function of the magnetic flux for the armchair nanotube; the figures \ref{zig-zagg}a
and \ref{zig-zagg}b show the same results but for the zig-zag nanotube.

The electronic structure of carbon nanotubes under uniform magnetic field
parallel to the tube axis was already study in the past using $k.p$
perturbation theory \cite{ando} and exact calculations \cite{lu}. A
magnetic field induced metal-insulator transition was then predicted: a
semiconductor nanotube becomes metallic for high enough magnetic field and,
reversely, a metallic nanotube becomes semiconductor. This dramatic behaviour
predicted theoretically could be an explanation for magnetoresistance
experiments on carbon nanotube bundles \cite{magnetoBundle} and more recent
ones on multi-wall carbon nanotubes \cite{magnetoMWCT}.

We recover these results in our calculations. The magnetic field opens a gap in
the case of  the armchair nanotube (figure \ref{armchairg}a); on the contrary, the magnetic
field tends to close the gap for the zig-zag nanotube until $\phi \simeq 0.35$
where the gap starts to increase smoothly (figure \ref{zig-zagg}a). These different
behaviours are also apparent in the polarizability as can be seen on the
figures \ref{armchairg}b and \ref{zig-zagg}b; the response functions follow the evolution of the band
gaps in both cases. Additionally, one can notice the peak observed at $\phi=0$ for the
armchair nanotube showing that the ground state and the first excited states
are directly coupled via the electric-field given rise to a strong linear Stark
effect.

The static electric polarizability tensor (without magnetic field) of carbon
nanotubes, $\vec{\alpha}$, was studied in the past for the half-filled case,
using a tight-binding model \cite{louie}. It was shown that the $\alpha_{zz}$
component of the polarizability tensor is proportional to $R/E_g^2$, where
$E_g$ is the band gap and $R$ is the radius of the tube, while $\alpha_{xx}$
is independent of $E_g$ and is proportional to $R^2$. In our case we are
concerned with $\alpha_{xx}$ and we have check the above mentioned scaling
law for zig-zag and armchair nanotubes. Our results, for these particular
achiral examples, are consistent with the study in \cite{louie}.

The $\alpha_{xx}$ component was studied in \cite{fulde} with an applied magnetic field
but using a semi-classical approximation which do not alow us to do direct
comparison with the results of the present work. However, it is reasonable to
think the absolute values of the polarizability tensor will not change
drastically by applying a magnetic field. Therefore, according to the results
of \cite{louie}, one can conclude that the static-magneto polarizability
should be much more intense for longitudinal electric field than transversal electric field.

The second results we want to show, as an illustration, are for the two very same systems but
slightly away from half-filling ($N_e=49000$). The corresponding induced
dipole moment are shown in figures \ref{nothalf}a and \ref{nothalf}b, for armchair and zig-zag
nanotubes, respectively, at low magnetic field only. Without going into any details
one immediately sees that both responses are considerably much intricate
than the ones at half-filling, indicating more complicated behaviours of the ground
and first-excited states as function of the magnetic field. The analysis of
such responses should give important informations about the electronic spectrum
of these compounds.

\section{Perturbative results}
In the linear regime, where we mainly worked, a perturbative expression for the
induced dipole moment should be appropriate. Let us consider a system
described by the general Hamiltonian $\hat{H}=\hat{H}_0+\hat{V}$ where
$\hat{H}_0$ is for the system without electric field and $\hat{V}$ takes
into account the effect of the electric field acting as a perturbation. At
second order in perturbation theory, the induced dipole moment is given by
\begin{equation}
D(T,\phi)=\frac{1}{2}\sum_{I,J}\frac{|<\Psi_{I}|\hat{V}|\Psi_{J}>|^2}{\epsilon_{I}-\epsilon_{J}}f_F(\epsilon_{I})(1-f_F(\epsilon_{J}))
\end{equation} where $\hat{H}_0 |\Psi_I>=\epsilon_I |\Psi_I>$ and
$f_F(\epsilon)=1/(e^{\beta(\epsilon-\mu)}+1)$ is the Fermi distribution
function, $\mu$ being the chemical potential.

Let us illustrate the effectiveness of these perturbative calculations for the
particular case of the cylinders of the section II, where the selection
rules are particularly restrictive. Indeed, in this case $\hat{V}$ is given by the
dipolar operator (\ref{dipoleoperator}) and the wave function, $| \Psi_I>$, by
Bloch functions (\ref{wavefunction}), $| \Psi_{p,q}>$. The matrix
elements of the dipole operator are
then given by the equation (\ref{selectionrules}), resulting in a simple
expression for the polarizability of a cylinder

\begin{equation}
\label{perturbation}
D(T,\phi)=\frac{e^2R^2}{2M}\sum_{p=0}^{N-1}\sum_{q=1}^{M} f_F(\epsilon_{p,q})\left
\{\frac{1-f_F(\epsilon_{p-1,q})}{\epsilon_p(\phi)-\epsilon_{p-1}(\phi)}+\frac{1-f_F(\epsilon_{p+1,q})}{\epsilon_p(\phi)-\epsilon_{p+1}(\phi)}
\right \}
\end{equation}where we add the dependence over the magnetic flux $\phi$. $R$
is the radius of the cylinder, $M$ and $N$ the number of sites along the cylindrical
axis and along the circumference, respectively, $\epsilon_{p,q}$ the spectrum defined in (\ref{spectrum}) and $\epsilon_p(\phi)=2t\cos(\frac{2\pi}{N}(p+\phi))$.

We have compared this perturbative expression with exact calculations for
different cylinders and several choices of temperature and values for the
hopping integrals, $t$ and $t_p$. The results are always in perfect
accordance, except for the very particular points where linear Stark effect occur, as
far as we remain in the linear regime. In this regime all the curves presented
in this work could have been obtained by the perturbative expression
(\ref{perturbation}). This could give a simplified framework to perform in the
future more sophisticated analysis of the problem like inclusion of electron-electron
interaction, for instance.

For nanotubes, more complex expressions will result due to less restrictive
selection rules. However, such perturbative calculations can also be done.

\section{Conclusions}

The main purpose of this work was to demonstrate, using exact
calculations, the strong magnetic field dependence of the static
polarizability of cylindrical systems when the magnetic field is parallel to the
tube axis. This was already predicted in \cite{fulde} on the basis of semi-classical
analysis. 

The demonstration was done for two kinds of system. On one hand we have
considered rolled square lattices and, on the other hand, two kinds of
non-chiral carbon nanotubes, i.e., metallic armchair nanotubes and the
zig-zag nanotubes, which can be either semiconducting or metallic \cite{saito-book}. For all these cases, the polarizability
was shown to present very complex structures as function of the magnetic field
in which one can identify two different
characteristics (cf figure 2): (i) the polarizability is a non-continuous function with
sudden jumps separating plateau-like regions (ii) additionally, small peaks
may appear for special values of the magnetic field in place of jumps.

A full understanding of these complicated behaviours was given by following
the behaviour of the ground state by increasing the magnetic field:
due to the Aharonov-Bohm effect, many changes of ground state occurs and for
some values of the magnetic field accidental degeneracies happen where the
ground state becomes two fold degenerate. A one to one correspondence is found
between the accidents in the polarizability and the accidental degeneracies of
the ground state. Each plateau-like region of the polarizability corresponds
to a quadratic Stark effect with a coefficient proper to the corresponding
magnetic-field induced ground state. Each peak corresponds to a linear Stark
effect appearing at accidental degeneracies when there is direct coupling
between the two states involved (cf figure 2). Therefore, it seems possible to
study the static polarizability under magnetic field in view to obtain
informations about excited states of cylindrical systems.

For ring shape cylinders and with a one-electron picture, the peaks due to
linear Stark effect appear for very particular magnetic field values,
$\phi=0.1/2,1$. The situation could be very different for different
shape - as we have seen for elliptical tubes - or with Coulomb interaction.

All the results shown in this work are for selected cylindrical
systems. Indeed, since the proposed measurements are extremely sensitive to the
characteristic sizes of the systems, $N$ and $M$, and to the electron density,
it is necessary to be able to select with high accuracy an individual
system. However, it is already possible to perform measurements on
individual single-wall nanotube \cite{selectednanotube}, for instance, which
make us to believe that the proposed experiments are nowaday possible to
realize.

Finally, we have done our studies with a one electron picture but the screening
due to electron-electron interaction is very important and can diminish
considerably the absolute value of $D(T,0)$ \cite{louie}. However, it is
reasonable to believe that the screening effects should be independent of the
applied magnetic field. Therefore, the ratio $\frac{D(T,\phi)-D(T,0)}{D(T,0)}$
should remain unchanged with screening effects.

Several extensions of this work are necessary. For the near future we are
planning to work in three directions. (i) We plan to consider instead of
individual, a set of cylindrical systems - with a
particular attention for set of nanotubes and multi-wall carbon nanotubes. (ii) An explicit treatment of
the electron-electron interaction is absolutely needed especially since important
qualitative changes could occur for ring systems \cite{kusmartsev} and very
important effects were shown on transport measurements of single-wall
nanotubes \cite{tans}. (iii) Disorder effects (topological or substitutional
disorder) are also of importance for the properties we are interested in
\cite{roche} and should be considered in the future.

\begin{acknowledgements}
It is a pleasure for us to thank Prof. P. Fulde for his support and careful
reading of the manuscript.
\end{acknowledgements}

\begin{figure}
\centerline{\psfig{figure=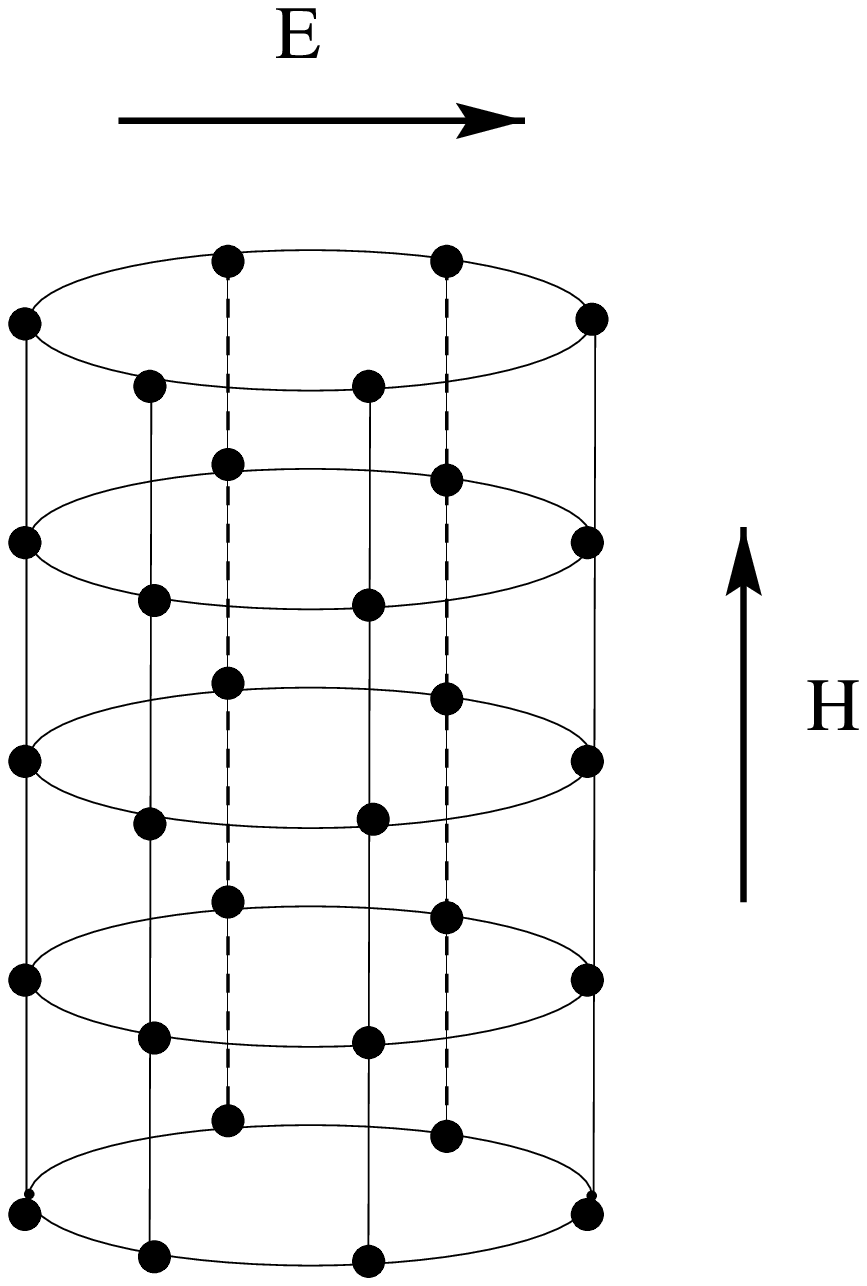,width=10cm}}
\caption{Cylinder made of rolled square lattice placed in two uniform
fields: an electric-field, $E$, perpendicular to the cylindrical axis and a
magnetic field, $H$, parallel to it. The same field configuration is adopted
for carbon nanotubes (rolled honeycomb lattices).}
\label{configuration}
\end{figure}

\begin{figure}
\centerline{\psfig{figure=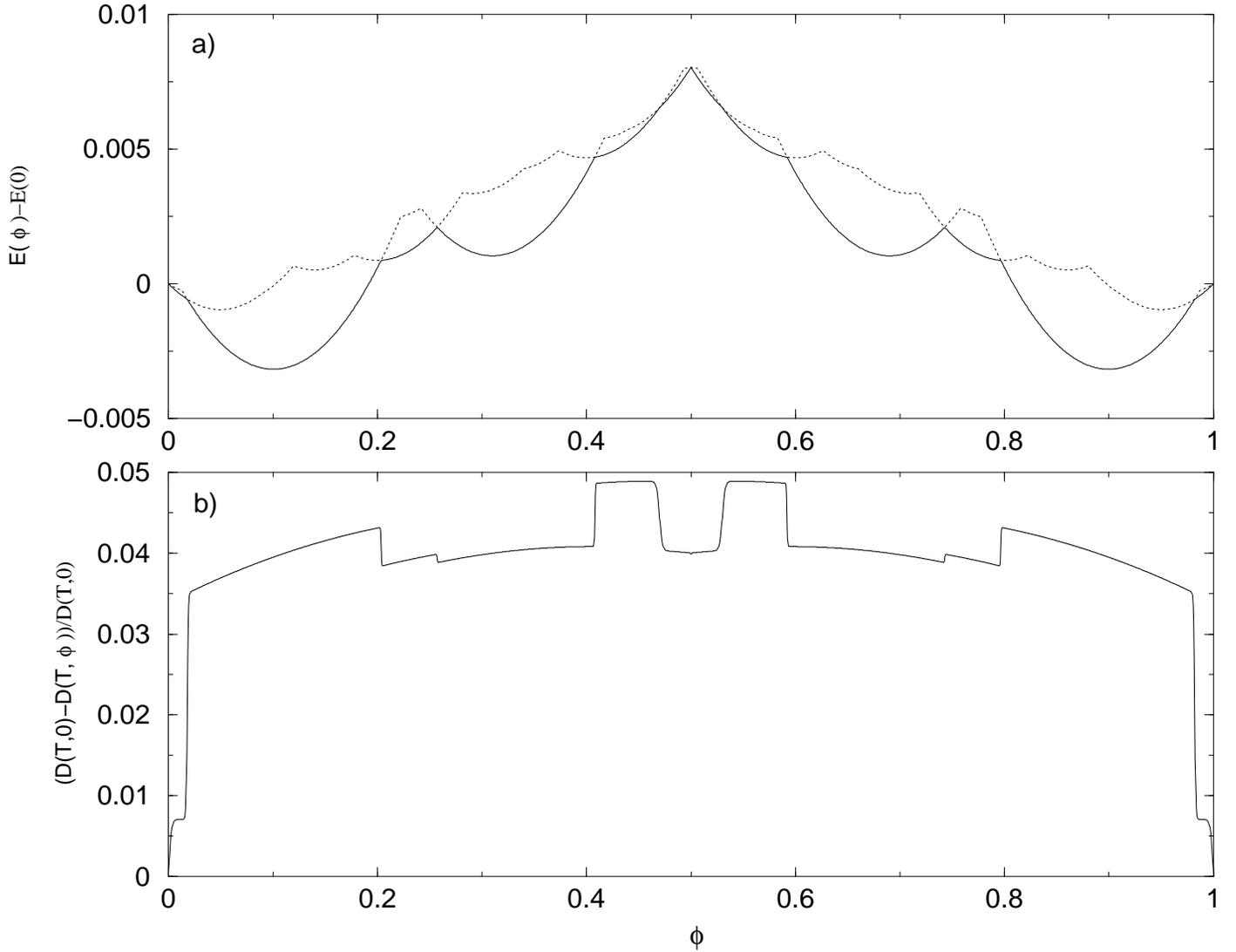,width=18cm,angle=-90}}
\caption{Cylinder with $N=101$, $M=100$ and $N_e=100$, the numbers of site
along the circumference, along the cylindrical axis and number of electrons
(respectively).(a) Energies of the ground state (full line) and of the first
excited state (dotted line) as function of the magnetic flux, $\phi$, without electric
field (the references are the energies without magnetic field). (b) Polarizability $(D(T,0)-D(T,\phi))/D(T,0)$, at $k_BT=10^{-5}t$ and for $v=eER=10^{-3}t$ as
function of the magnetic flux; $t$ is the hopping integral defined in
(\ref{huckel}), $E$ the electric field and $R$ the radius of the cylinder.}
\label{characteristics}
\end{figure}

\begin{figure}
\centerline{\psfig{figure=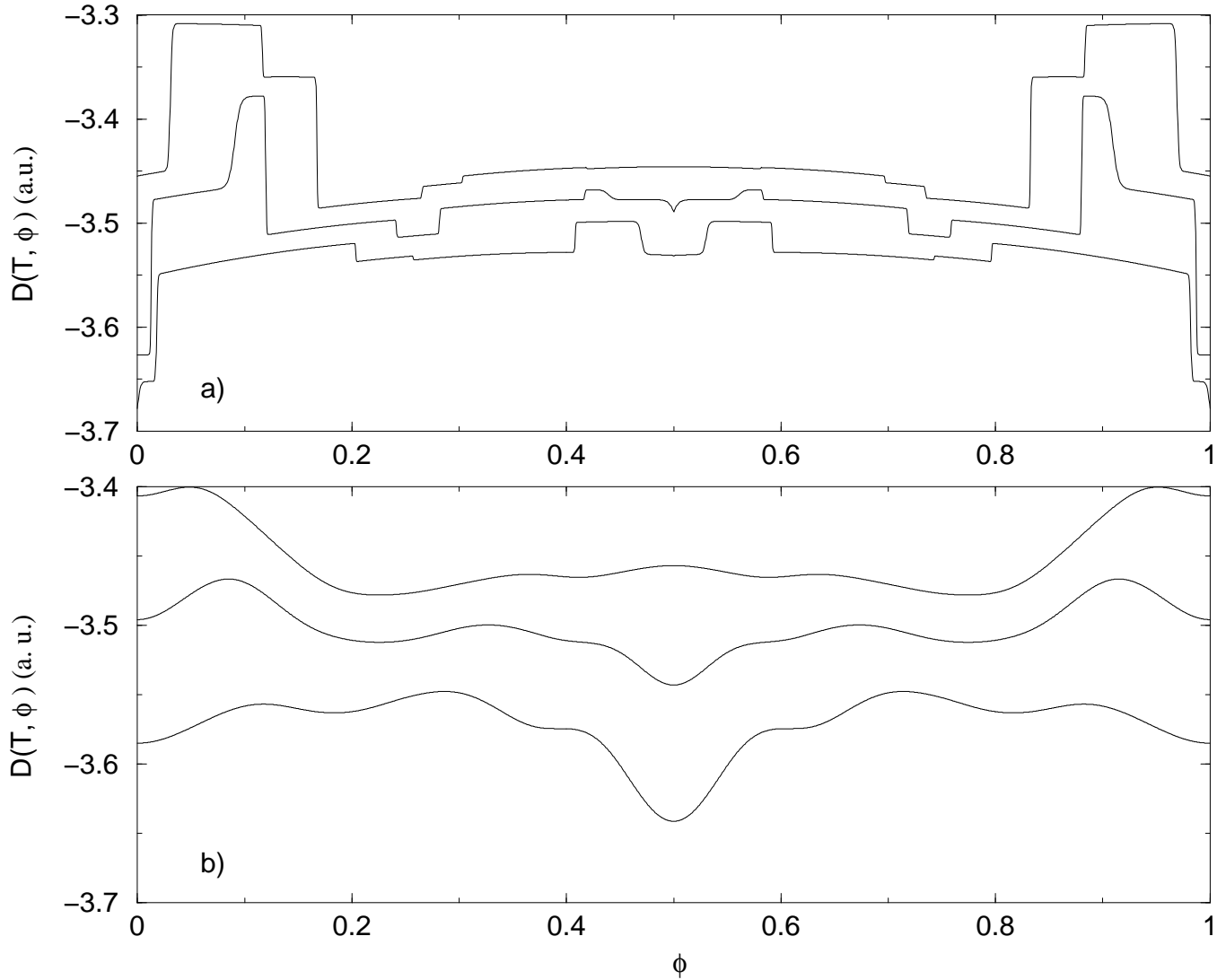,width=18cm,angle=-90}}
\caption{Polarizability of a cylinder with $N=101$ and  $M=100$ as function of
the magnetic flux $\phi$ in arbitrary units (a.u.). (a) For $N_e=100, 101, 102$ from the bottom to the top at a
temperature of $k_BT \simeq 10^{-1}K$. (b) The same but for $k_BT \simeq 10K$.}
\label{fillingeffects}
\end{figure}

\begin{figure}
\centerline{\psfig{figure=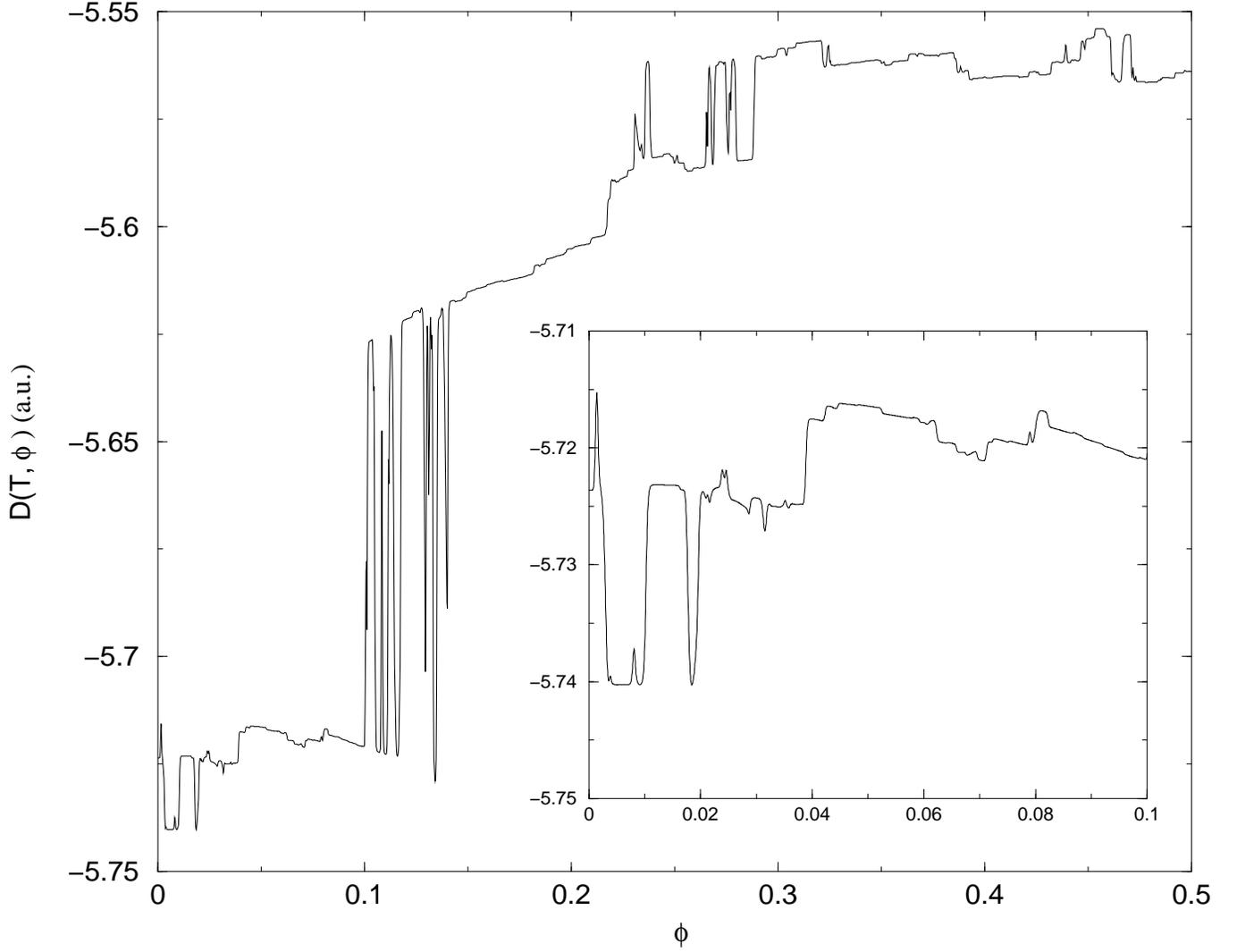,width=18cm,angle=-90}}
\caption{Polarizability of a cylinder with $N=101$, $M=1000$ and $N_e=20000$
as function of the magnetic flux at $k_BT=10^{-5}t$ in arbitrary units (a.u.). In the inset: zoom of the small magnetic field part.}
\label{complexcylinder}
\end{figure}

\begin{figure}
\centerline{\psfig{figure=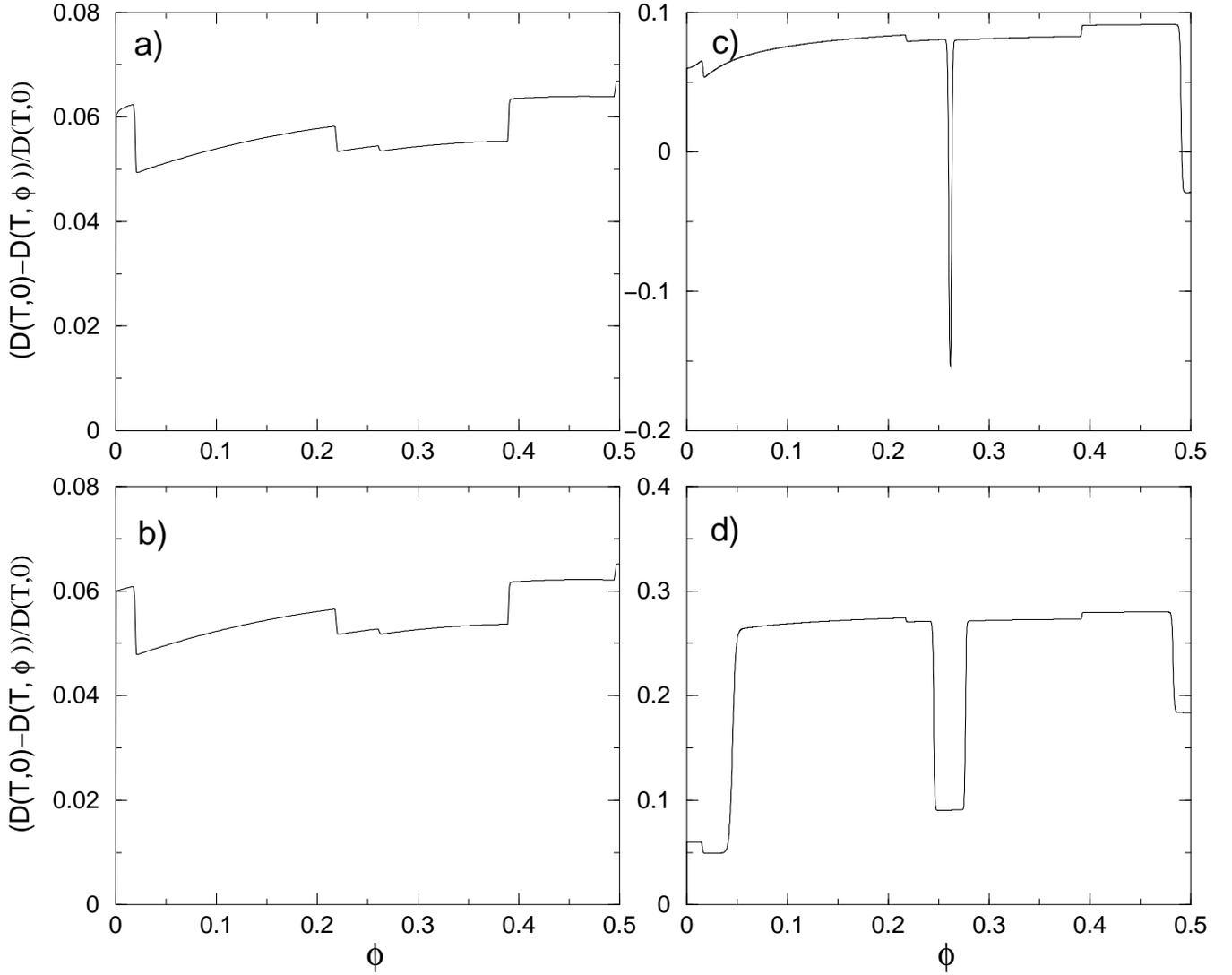,width=18cm}}
\caption{Polarizability $(D(T,0)-D(T,\phi))/D(T,0)$ of an ellipse and a cylinder with $N=100$, $M=100$ and
$N_e=100$ as function of the magnetic flux, $\phi$, at $k_BT=10^{-5}t$. (a) Ellipse with
$\frac{a^2}{b^2}=0.25$ and $v=10^{-3}t$. (b) Cylinder with $v=10^{-3}t$. (c)
Ellipse with $\frac{a^2}{b^2}=0.25$ and $v=10^{-2}t$. (d) Cylinder with $v=10^{-2}t$.}
\label{ellipse}
\end{figure}

\begin{figure}
\centerline{\psfig{figure=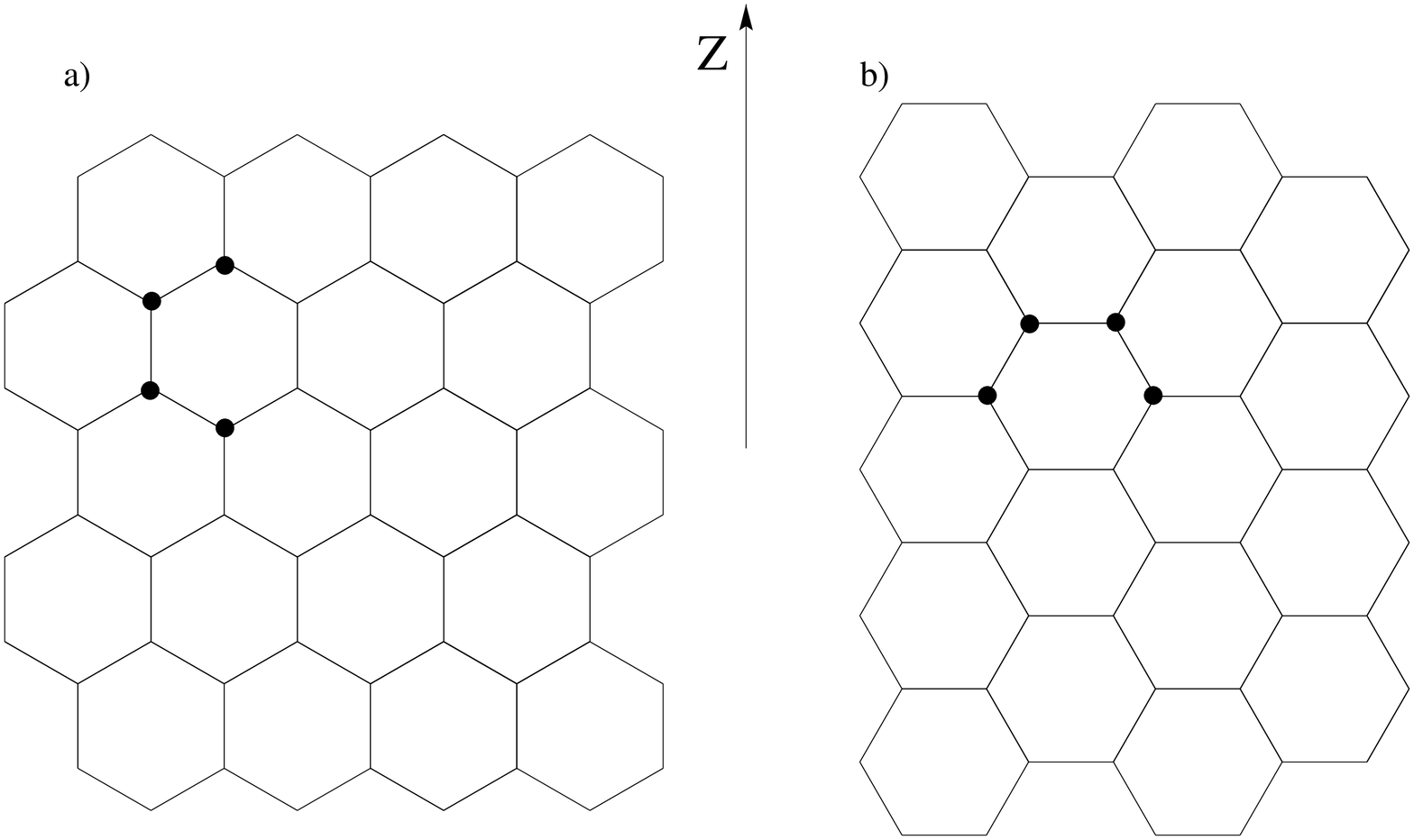,width=18cm}}
\caption{Carbon nanotubes are rolled honeycomb lattices around the
Z-axis. Here are represented part of the honeycomb lattice for (a) Zig-Zag
carbon-nanotubes and (b) Armchair carbon-nanotubes. They are both achiral
nanotubes, similar to cylindrical systems with 4 carbon atoms per unit cell
marked here by the black dots.}
\label{nanolattice}
\end{figure}

\begin{figure}
\centerline{\psfig{figure=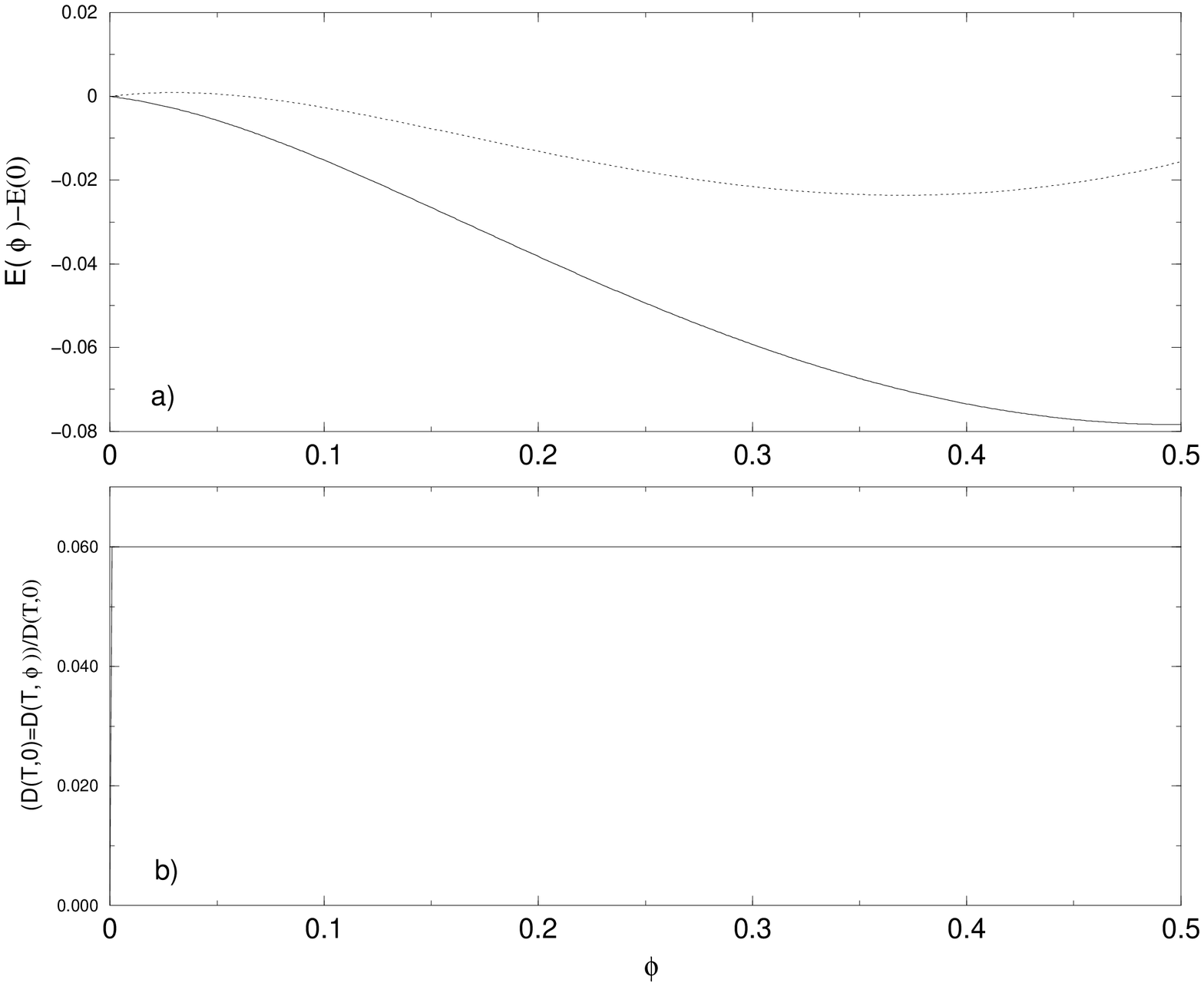,width=18cm,angle=-90}}
\caption{Armchair nanotube with $N=50$, $M=500$ and $N_e=50000$ (half-filling)
at $k_BT=10^{-5}t$. (a) Energies of the ground and first excited state as
function of the magnetic flux $\phi$ (the references are the energies without
magnetic field). (b) Polarizability $(D(T,0)-D(T,\phi))/D(T,0)$ for $v=10^{-3}t$ as function of the magnetic flux.}
\label{armchairg}
\end{figure}

\begin{figure}
\centerline{\psfig{figure=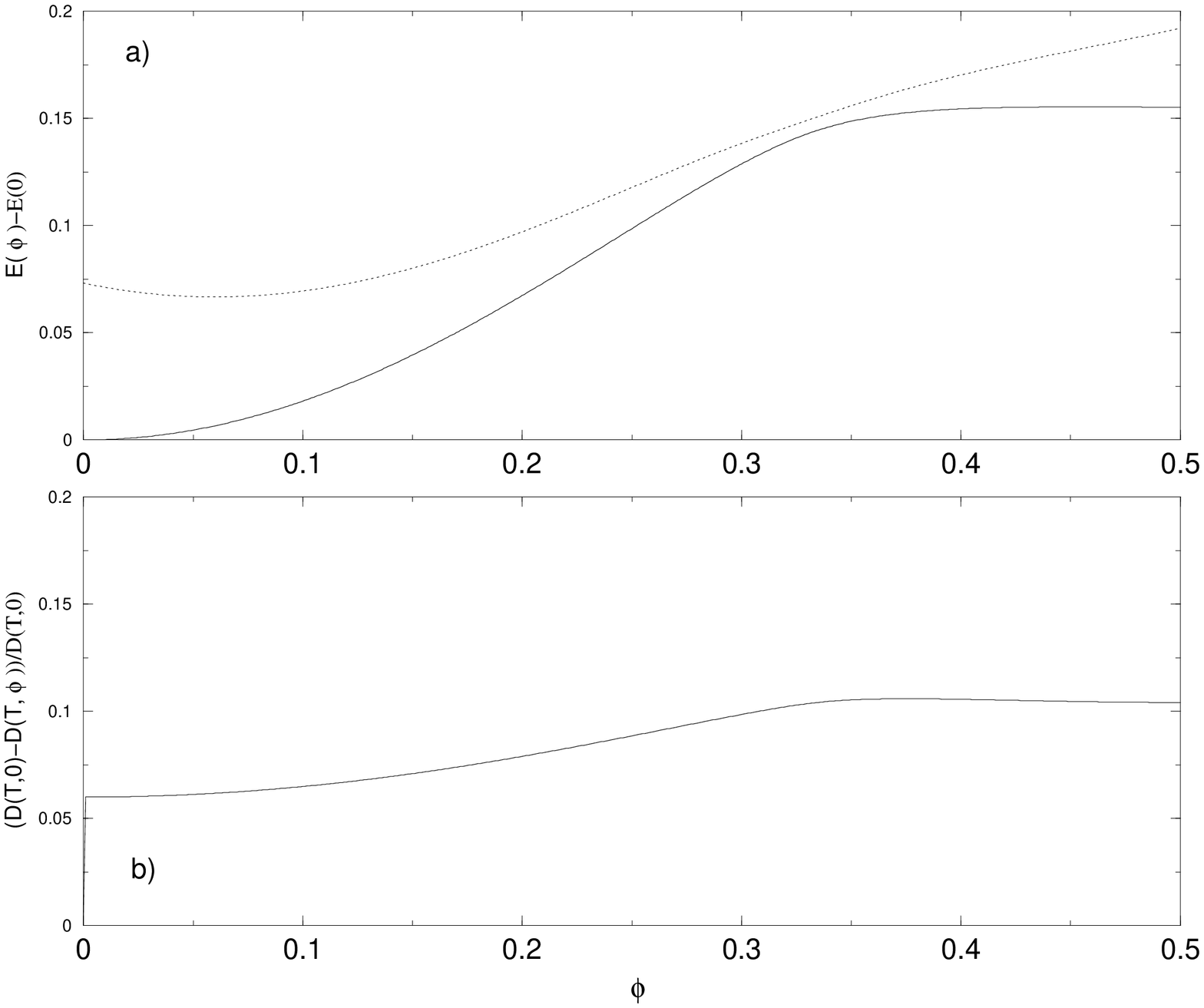,width=18cm,angle=-90}}
\caption{Zig-Zag nanotube with $N=50$, $M=500$ and $N_e=50000$ (half-filling)
at $k_BT=10^{-5}t$. (a) Energies of the ground and first excited state as
function of the magnetic flux $\phi$ (the references are the energies without
magnetic field). (b) Polarizability $(D(T,0)-D(T,\phi))/D(T,0)$ for $v=10^{-3}t$ as function of the magnetic flux $\phi$.}
\label{zig-zagg}
\end{figure}

\begin{figure}
\centerline{\psfig{figure=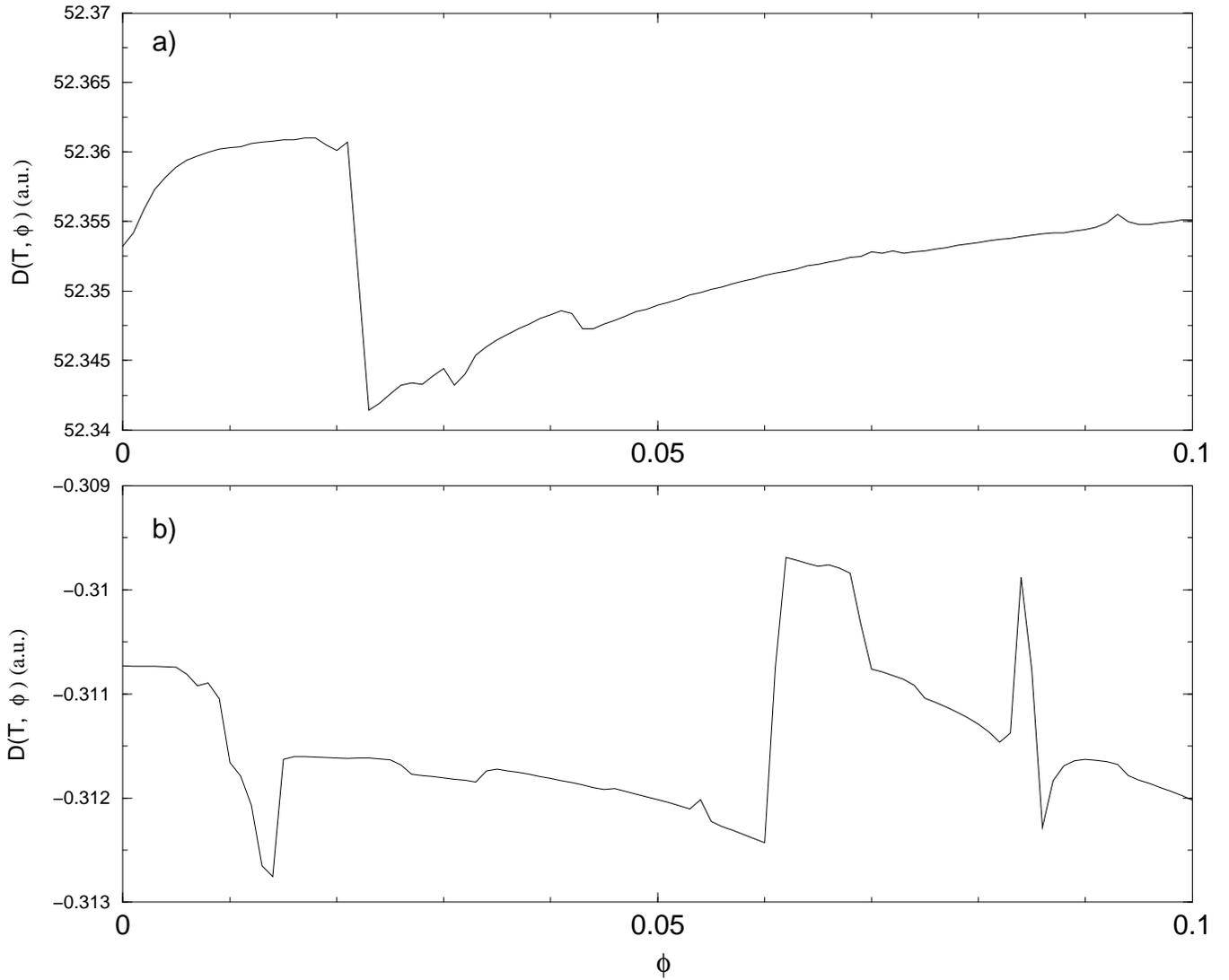,width=18cm,angle=-90}}
\caption{Polarizability in arbitrary units (a.u.) of (a) armchair and (b) zig-zag nanotubes with $N=50$,
$M=500$, $N_e=49000$, $k_BT=10^{-5}t$ and $v=10^{-3}t$ as function of the magnetic flux $\phi$.}
\label{nothalf}
\end{figure}

\end{document}